\documentclass[prd,floats,aps,epsfig,nofootinbib,axodraw,amssymb,preprintnumbers]{revtex4}
\usepackage{amsmath,amssymb,amsfonts}
\usepackage{graphicx,epsfig,url}
\usepackage{color}
\usepackage{amsmath}
\usepackage{amsfonts}
\usepackage{amssymb}
\usepackage{slashed}
\usepackage{graphicx}
\usepackage{tikz}
\usetikzlibrary{positioning}
\usetikzlibrary{decorations.pathmorphing}
\usetikzlibrary{decorations.markings}
\usetikzlibrary{snakes}
\usepackage{float}
\usepackage{caption}
\usepackage{graphicx}%
\numberwithin{equation}{section} 
\setcaptionmargin{1cm}
\input{epsf}
\def\beq{\begin{equation}}
\def\eeq{\end{equation}}
\def\bea{\begin{eqnarray}}
\def\eea{\end{eqnarray}}
\def\bitem{\begin{itemize}}
\def\eitem{\end{itemize}}
\def\leqn#1{(\ref{#1})}
\def\bar#1{\overline{#1}}

\def\inv{^{\raise.15ex\hbox{${\scriptscriptstyle -}$}\kern-.05em 1}}
\def\lbar{{\lower.35ex\hbox{$\mathchar'26$}\mkern-10mu\lambda}} 

\def\pslash{\not{\hbox{\kern-4pt $p$}}}
\def\qslash{\not{\hbox{\kern-4pt $q$}}}
\def\lv{\not{\hbox{\kern-4pt $L$}}}
\def\lsim{\mathrel{\raise.3ex\hbox{$<$\kern-.75em\lower1ex\hbox{$\sim$}}}}
\def\gsim{\mathrel{\raise.3ex\hbox{$>$\kern-.75em\lower1ex\hbox{$\sim$}}}}
\def\ifmath#1{\relax\ifmmode #1\else $#1$\fi}

\def\to{\rightarrow}

\let\<=\langle
\let\>=\rangle

\let\+=\uparrow

\begin{document}
\tikzset{ 
  scalar/.style={dashed},
  scalar-ch/.style={dashed,postaction={decorate},decoration={markings,mark=at
      position .5 with {\arrow{>}}}},
  fermion/.style={postaction={decorate}, decoration={markings,mark=at
      position .5 with {\arrow{>}}}},
  gauge/.style={decorate, decoration={snake,segment length=0.2cm}},
  gauge-na/.style={decorate, decoration={coil,amplitude=4pt, segment
      length=5pt}}
}
\newcommand{\JB}[1]{{\color{red} JB: #1}}
\begin{flushright}                                                                                                                                                                                                                                                                                                                                                                                                                                                                                                                                                                                                                                             SLAC-PUB-16533
\end{flushright}

\title{Cosmological Constraints on Decoupled Dark Photons and Dark Higgs}
\bigskip
\author{Joshua Berger$^a$, Karsten Jedamzik$^b$ and Devin~G.~E.~Walker$^c$ \vspace{0.02in}}
\address{$^a$Department of Physics, University of Wisconsin-Madison, Madison, WI 53706, USA}
\address{$^b$Laboratoire Univers et Particules de Montpellier, UMR5299-CNRS, Universit\'{e} Montpellier II, 34095 Montpellier, France}
\address{$^c$Department of Physics, University of Washington, Seattle, WA 98195, USA}

\begin{abstract}
\noindent
Any neutral boson such as a dark photon or dark Higgs that is part of a non-standard sector of particles can mix with its
standard model counterpart. When very weakly mixed with the Standard Model, these particles are produced in the early Universe via the freeze-in mechanism and subsequently decay back to standard model particles.  In this work, we place constraints on such mediator decays by considering bounds from Big Bang nucleosynthesis and the cosmic microwave background radiation. 
We find both nucleosynthesis and CMB can constrain dark photons with a kinetic mixing
parameter between $\log \epsilon \sim$ -10 to -17 for masses between 1
MeV and 100 GeV.  Similarly, the dark Higgs mixing angle $\epsilon$ with the Standard
Model Higgs is constrained between $\log \epsilon \sim$ -6 to -15.  %
Dramatic improvement on the bounds from CMB spectral distortions can be achieved with
proposed experiments such as PIXIE.
\end{abstract}
\maketitle

\tableofcontents

\section{Introduction}
\label{sec:intro}
\noindent
Cosmological evidence indicates that the majority of matter in the
universe is non-baryonic dark matter.  Given its cosmological
importance, there is little reason to believe that dark matter is part
of some simple, inert sector.  The interactions of the dark sector
could be as complex as those of our own visible sector.  Bosons that
mediate dark matter self-interactions could also form a  portal by which
the SM can interact weakly with dark matter.  In such a scenario, a
neutral gauge/scalar boson meditor mixes with the SM counterpart
photon-Z/Higgs systems.  The neutral gauge boson particles are
referred to as dark photons
\cite{Holdom:1985ag,DelAguila:1993px,DelAguila:1995fa,Babu:1996vt,Babu:1997st}
(even though they also mix in part with the $Z$ boson), while the neutral
scalar boson particles are referred to as dark Higgs
\cite{Silveira:1985rk,Foot:1991bp,Foot:1991py,Chacko:2005pe,Barbieri:2005ri,Chang:2006fp,Barger:2007im,Strassler:2006ri,Strassler:2006im,Lebedev:2011aq,Chang:2007ki,Englert:2011yb,Wells:2008xg,Schabinger:2005ei,Bowen:2007ia,Patt:2006fw}.
These interactions could well dominate the interaction between the dark and
visible sectors as they are among the only possible renormalizable
interactions of a dark sector with the SM.
\newline
\newline
As such, portal interactions have been studied extensively \cite{Englert:2013gz,Kumar:2012ww,Englert:2012ha,LopezHonorez:2012kv,Lebedev:2012zw,Batell:2011pz,Djouadi:2011aa,Englert:2011aa,Baek:2011aa,Lebedev:2011iq,Brivio:2015kia,Sun:2015oea,Freitas:2015hsa,Fedderke:2015txa,Khoze:2015sra,Bishara:2015cha,Chao:2015uoa,Falkowski:2015iwa,Chao:2014ina,Chacko:2013lna,Choi:2013qra}.
Much of the focus has been on the region of relatively large mixing,
wherein the mediator itself is in thermal equilibrium with the SM bath
in the early universe before rapidly decaying when $H < \Gamma$.
Constraints on mediators in such scenarios arise from a combination of
laboratory experiments such as beam dumps and lepton colliders, as
well as astrophysical constraints from supernovae for lighter
mediators
\cite{Lees:2014xha,Blumlein:2011mv,Andreas:2012mt,Endo:2012hp,Babusci:2012cr,Babusci:2014sta,Adlarson:2013eza,Agakishiev:2013fwl,Blumlein:2013cua,Merkel:2014avp,Abrahamyan:2011gv,Lees:2012ra,Aubert:2009af,Aubert:2009cp,Kazanas:2014mca,Batell:2014mga,Anastasi:2015qla,Holzmann:2014eya,Khachatryan:2015wka,Aad:2015sms,Batley:2015lha,Babusci:2015zda,TheBelle:2015mwa,Palladino:2015jya,Curciarello:2015xja,Adare:2014mgk,Aad:2014yea}.
Their direct effect on cosmology is negligible
since they decay well before the epochs we can study observationally,
i.e.\ Big Bang Nucleosynthesis (BBN) and recombination.  On the other hand, for
smaller mixing angles, the mediator can be sufficiently long lived to
decay during or after one of these events, opening the door for a new
set of constraints that restrict mediator parameter space at small
mixing angles.
\newline
\newline
In order for the mediators to have a significant effect, they must be
produced in non-neglible quantities.  This is however the case, since even under
the pessimistic assumption that no mediators were present in the very
early universe, such as the period just after inflaton decay, a
sufficient abundance will be ``frozen-in'' by mixing with the SM
bosons so as to constrain much of the parameter space.  This process
of mediator freeze-in could also lead to the generation of a
significant abundance of dark matter, though we leave the study of
such an effect to future work.
\newline
\newline
In this work we characterize the effect of the mediator decays on
nucleosynthesis and the CMB.  We show, even in
the limit where the mediators have a small coupling with the Standard
Model, large regions of parameter space are excluded.  We note, however, that our
constraints do evaporate in case these dark mediators decay into lighter stable dark 
sector particles before decaying into standard model particles. Constraints on 
weakly mixed dark photons with
Stueckelberg masses were published previously in Ref.~\cite{Fradette:2014sza}.
Our analysis signifcantly improves on their study by correcting some errors and
negligence in their BBN treatment.
\newline
\newline
The remainder of this paper is structured as follows.  In Section
\ref{sec:model}, we review the dark photon and dark Higgs models.
Subsequently, we study the production and decay of mediators in
Sections \ref{sec:prod} and \ref{sec:mediator-decays}.  We then
present constraints on both models in Sections \ref{sec:bbn} and
\ref{sec:cmb}.  Finally, we conclude in Section
\ref{sec:conclusions}.

\section{Mediator Models}
\label{sec:model}   
\noindent
We consider a generic scenario where the Standard Model (SM) gauge
group is extended by an additional $U(1)_D$ which we refer to as
dark hypercharge. A dark Higgs is introduced to spontaneously break
$U(1)_D$ and give mass to the corresponding dark photon.
%
\subsection{Dark Higgs Mixing and Couplings}
\noindent
In the minimal model, an additional scalar boson $\Phi$ charged under a new
$U(1)_D$ gauge group is introduced.  The scalar boson gets a vacuum
expectation value that breaks $U(1)_D$, giving mass to the dark
Higgs. We parametrize the Higgs fields as
\begin{align}
  H = {1 \over \sqrt{2}}\,\begin{pmatrix}
    0\\
    v + h 
 \end{pmatrix}
 && \Phi = {1 \over \sqrt{2}}\bigl(u + \rho \bigr)\, \label{eq:vevs},
 \end{align}
neglecting the Goldstone modes which get eaten.
While we shortly consider potential kinetic mixing between the $U(1)_D$
and $U(1)_Y$ gauge bosons, we first consider mixing between $\rho$ and the
SM Higgs field $H$.  The most general scalar potential after symmetry
breaking can be written as
\begin{align}
V &= \lambda_1 \biggl( H^\dagger H - \frac{v^2}{2}\biggr)^2 + \lambda_2 \biggl( \Phi^\dagger\Phi - \frac{u^2}{2}\biggr)^2 + \lambda_3\biggl( H^\dagger H - \frac{v^2}{2}\biggr)\biggl( \Phi^\dagger\Phi - \frac{u^2}{2}\biggr) \, . \label{eq:higgspotential}
\end{align}
The $\lambda_3$ coupling mixes the visible and dark Higgs sectors.
The physical Higgs masses are
\begin{align}
m_h^2 = 2\,\lambda_1 v^2\biggl(1 - {\lambda_3^2 \over 4\,\lambda_1\lambda_2} + \ldots\biggr) 
&& m_\rho^2 = 2\,\lambda_2\,u^2\biggl(1  +{\lambda_3^2 \over 4\,\lambda_2^2} {v^2\over u^2}+ \ldots\biggr) \label{eq:Higgsmass}
\end{align}
where $m_h$ is the SM Higgs mass and is fixed to 125
GeV~\cite{Aad:2015zhl}.  $m_\rho$ is the dark Higgs mass.  The Higgses
will mix as
\begin{equation}
\begin{pmatrix} h' \\ \rho' \end{pmatrix} = \begin{pmatrix} 
\cos\epsilon & \sin\epsilon \\ -\sin\epsilon & \cos\epsilon \end{pmatrix}\, \begin{pmatrix} h \\ \rho \end{pmatrix}, \label{eq:mixing}  
\end{equation}
where the primes denote the mass eigenstates.  For brevity going
forward, we refer to both the mass eigenstates without primes.  The
mixing angle is given by
\begin{equation}
  \tan 2\epsilon = \frac{\lambda_3 u v}{\lambda_1 v^2 - \lambda_2 u^2}
\end{equation}
In the limit where $\lambda_3 u v \ll \lambda_1 v^2,\lambda_2 u^2$,
which is the limit considered in this work, the mixing angle can be
written as
\begin{equation}
  \label{eq:14}
  \tan 2\epsilon = \frac{2 \lambda_3 u v}{m_h^2 - m_\rho^2}.
\end{equation}
Given equation~\leqn{eq:mixing}, the SM fermion coupling to the mass
eigenstate dark Higgs is simply a rescaling of the SM coupling,
\begin{equation}
y_{\bar{f}f\rho} = -{i g \,m_f \over 2 \,m_W} \sin\epsilon\,.
\end{equation}
There are enough free parameters to treat the dark Higgs mass and
mixing parameters as independent parameters and we do so in this work.
The corresponding coupling between the SM Higgs and the SM 
fermions is proportional to $\cos\epsilon$.  Because we are interested 
in the regime of parameter space where $\sin\epsilon$ is very small,
constraints on SM Higgs couplings from the LHC are not
applicable~\cite{Aad:2015gba,Khachatryan:2014jba}.

Note that there are actually 3 new parameters in this model, which we
can take to be either $\lambda_2$, $\lambda_3$ and $u$, or more
conveniently $m_\rho$, $\epsilon$, and $\lambda_3$.  While most of the dark
Higgs couplings to the SM depends only on $\epsilon$ and kinematics
depends only on $m_\rho$, the $\rho^2 h^2$ coupling
depends directly on $\lambda_3$.  We can consistently
take the limit $\lambda_3 \to 0$, holding $\epsilon$ fixed by
simultaneously taking $u \to \infty$.  In this work, we adopt such a
limit, which is the most conservative option: deviation from
this limit would enhance dark Higgs production, while keeping the
lifetime fixed.  The extra dark Higgs abundance can only enhance
constraints.  

\subsection{Dark Photon Mixing and Couplings}
\noindent
The dark Higgs also generates a mass for the dark photon. In this
section, we focus on the properties of this massive gauge boson.  For
concreteness, we take $Q_D = 2$ for the dark Higgs, 
though this choice is arbitary up
to a rescaling of the dark gauge coupling.  The gauge sector lagrangian is
\begin{equation}
\mathcal{L}_\mathrm{gauge} = -\frac{1}{4}\hat{B}_{\mu\nu}\hat{B}^{\mu\nu}-\frac{1}{4}\hat{F'}_{\mu\nu}\hat{F'}^{\,\mu\nu} - \frac{\epsilon}{2}\hat{B}_{\mu\nu}\hat{F'}^{\,\mu\nu} \, ,\label{eq:gaugewithkineticmix}
\end{equation}
where $\epsilon$ is the coupling of the kinetic term and links the
dark and visible $\rm U(1)_Y$ gauge sectors.  Because of this coupling, any SM
particle with non-zero hypercharge will then acquire a dark charge.
Because the $U(1)_D$ gauge boson mixes with hypercharge, the dark
photon will have chiral couplings to the SM fermions.  The axial
component, however, is suppressed in the limit that $m_{\gamma'} \ll m_Z$,
as we see explicitly below.  Without loss of generality, we parametrize this coupling as $\epsilon = \sin\delta$~\cite{Babu:1997st}.  
The kinetic terms can be diagonalized by defining new fields $B_\mu$
and $A'_\mu$ such that
\begin{align}
\begin{pmatrix}
\hat{B}_\mu \\
\hat{A'}_\mu
\end{pmatrix} = 
\begin{pmatrix}
1 & -\tan\delta\\
0 & \sec\delta 
\end{pmatrix}
\begin{pmatrix}
B_\mu \\
A'_\mu
\end{pmatrix}
\end{align}
where the hatted fields are the fields before diagonalizing the
kinetic mixing.  After this rotation,
equation~\leqn{eq:gaugewithkineticmix} is diagonalized.  However, a
mixing is induced in the mass matrix for the gauge bosons.   After
doing  the standard Weinberg angle diagonalization, we have 
\begin{eqnarray}
\left|D_\mu H\right|^2  + \left|D_\mu \phi\right|^2  &\supset&  {1 \over 2} m_Z^2 \,Z^2 +  {1 \over 2} m_{\gamma'}^2 \,A^{\prime\,2} +  \delta m^2 \,Z A'
\end{eqnarray}
where
\begin{eqnarray} 
m_Z^2 &=& {1 \over 4} v^2 \bigl( g_1^2 + g_2^2 \bigr) \label{eq:mz} \\ 
m_{\gamma'}^2 &=& {1 \over 4} v^2  g_1^2 \tan^2\delta + 4 \,u^2 \,g'^{\,2} \sec^2\delta \label{eq:mzp} \\
\delta m^2 &=& - {1 \over 4} v^2\, g_1 \sqrt{g^2_1 + g_2^2} \, \tan\delta\,\,.
\end{eqnarray}
We can diagonalize the mass term by introducing a mass mixing angle $\xi$ defined by
\begin{eqnarray}
\tan 2\xi 
= 
\biggl( {2 \,m_Z^2 \over m_Z^2 - m_{\gamma'}^2}  \biggr)\sin\theta_W \tan\delta \label{eq:massmix} \,.
\end{eqnarray}
In the limit of small kinetic mixing, the mass eigenstates are
approximately the charge eigenstate mass terms given in
equations~\leqn{eq:mz} and~\leqn{eq:mzp}.  The coupling between the SM
fermions, $f$, and the dark photon is
\begin{equation}
g_{\bar{f}f\,\gamma'} = -i\,\gamma^\mu\, \bigl(g'_{V} + g'_{A}\gamma_5 \bigr)\,. \label{eq:darkphoton-fermioncoupling}
\end{equation}
where 
\begin{align}
g'_{V} = - e \epsilon \left( Q \frac{c_W^2 m_Z^2 - m_{\gamma'}^2}{c_W (m_Z^2 - m_{\gamma'}^2)}
- T_3 \frac{m_{\gamma'}^2}{2 c_W (m_Z^2 - m_{\gamma'}^2)} \right)
&& g'_{A} = e \epsilon T_3 \frac{m_{\gamma'}^2}{2 c_W (m_Z^2 - m_{\gamma'}^2)}  \,
\end{align}
Here $c_W$ is the cosine of the Weinberg angle with $c_W^2\approx 0.775$.
Thus, for the SM fermion, $f$, the only unknown
parameter is the kinetic mixing parameterization.  In the limit
$m_{\gamma'} \ll m_Z$, this is proportional to the photon couplings to $f$:
\begin{align}
  g'_{V} \approx - c_W e Q \epsilon && g'_{A} \approx 0
\end{align}
  As in
the Higgs case, there are enough free parameters to treat the dark
photon mass and mixing as independent parameters.

\section{Freeze-In Production of Mediators}
\label{sec:prod}
\noindent
As emphasized above, we are considering scenarios in which the mixing
between the mediator and the SM is very small.  Thus, the rate at
which the mediator interacts with the SM is much less than the Hubble rate at
all times over the vast majority of parameter space\footnote{For some
  portions of the dark Higgs parameter space considered in this work,
  the mediator production becomes large and equilibrium is reached.
  However, we do not attempt to further study such portions of
  parameter space.}.  Even if the mediator never equilibrates with the
SM, it can be produced non-thermally and have observable effects on
cosmological evolution.  In the limit of such small mixing, the
mediators can be produced via two mechanisms: decay of heavy,
non-standard particles and freeze-in.  We adopt a conservative
approach and assume only a negligible amount of mediator particles are
produced by early decays of non-standard particles, including any
inflatons.  The abundance is then entirely determined by freeze-in
production mediated by SM particles~\cite{Hall:2009bx}.
In this section, we determine the freeze-in abundance of both dark
photons and dark Higgs.
\newline
\newline
Provided the coupling of the mediator to light particles is similar to
the coupling of the mediator to heavy particles, then mediator
production will be dominated by inverse decay processes.  Processes
with a higher number of initial states are phase space suppressed.
This criterion is in fact satisfied in the case of the dark photon,
where the coupling between all standard model particles 
and the mediator are comparable.  In the case of the dark
Higgs, however, the coupling to particles is proportional to their mass.  We find that the additional phase space suppression of $2 \to 2$ processes is more than overcome by the larger coupling to heavy particles and production is dominated by top quark annihilation and inelastic scattering.

\subsection{Dark Photon Freeze-In Production}
\label{sec:dark-phot-prod}
\noindent

\begin{figure}
\epsfxsize=9.5cm
\epsffile[50 50 410 302]{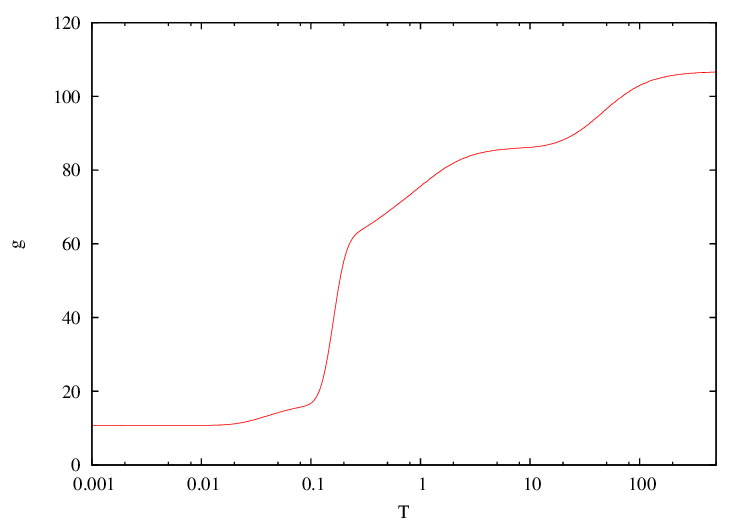}
\caption{The relativisitic degrees of freedom $g(T)$ as a function of temperature
T in GeV.} 
\label{fig:gs}
\end{figure}

The dominant production of dark photons comes from inverse decay of charged 
species.  We assume that $m_{\gamma'} <
2 \,m_W$, so that the decays are to charged fermions: quarks and leptons. 
The decay rate of the dark photon to species $\psi_i$ can be written as
\begin{equation}
  \label{eq:dp-decay}
  \Gamma_i \equiv \Gamma(\gamma' \to \psi_i \bar{\psi}_i) = \epsilon^2
  \frac{N_c c_W^2 Q_i^2 \alpha}{3} 
  m_{\gamma'}\biggl(1 - \frac{4m_i^2}{m_{\gamma'}}\biggr)^{1/2},
\end{equation}
where $N_c$ is the number of colors for species $i$. The freeze-in 
production of the dark photon is most easily computed by the principle of
detailed balance, i.e. the thermal rate for
inverse decay production equals that of the decay rate of the 
dark photon with a putative equilibrium distribution
\begin{equation}
\label{eq:2}
\frac{{\rm d} n_{\gamma'}}{{\rm d} t} = \int_0^{\infty}
\frac{{\rm d^3} p_{\gamma'}}{(2\pi)^3} 
\frac{f_{\gamma'}^{eq}\Gamma_{\gamma'}}{E_{\gamma'}/m_{\gamma'}}\, ,
\end{equation}
where
\begin{equation}
  \label{eq:2a}
f_{\gamma'}^{eq} = \frac{g_{\gamma'}}{e^{E_{\gamma'}/T} - 1}
\end{equation}
is the dark photon equilibrium distribution with $g_{\gamma'} = 3$ 
the dark photon number of degrees of freedom and where
$\Gamma_{\gamma'}=\sum_i\Gamma_i$ is the total width of the dark photon.

This expression may be used to compute the asymptotic dark-photon-to-entropy
ratio $Y_{\gamma^{'}} = n_{\gamma{'}}/s$, produced by inverse decay 
of fermions in the early Universe at temperatures $T {}^>_{\sim} m_{\gamma'}$
\begin{equation}
  \label{eq:4}
  Y(t \to \infty) \approx 
  \frac{g_{\gamma'}}{g_(m_{\gamma'})}
  \frac{90}{\pi^4}
  \frac{\Gamma_{\gamma'}}{H(m_{\gamma'})} \, ,
\end{equation}
where $H(T)$ is the Hubble scale at temperature $T$.
The largest contribution to the dark photon abundance comes from
temperatures of order the dark photon mass.  The dark photon abundance
$Y$ increases as $1 / T^3$ before reaching a maximum at $T \sim
m_{\gamma'}$. Eq.~\eqref{eq:4} is only correct up to a factor of order unity.
This is mostly due to an assumed constancy of the relativisitic degrees of 
freedom during the final phases of freeze-in. 
As may be seen from Fig.~\ref{fig:gs} the relativistic degrees of freedom change drastically
particularly during the QCD epoch. Results given below include are exact
as Eq.~\eqref{eq:2} is numerically integrated for the
analysis.

\subsection{Dark Higgs Freeze-In Production}
\label{sec:dark-higgs-abundance}
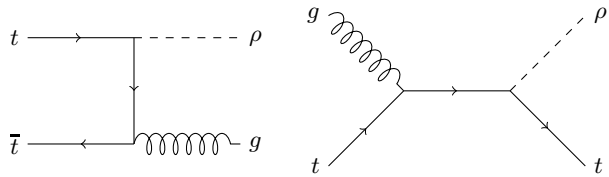
\begin{figure}[!tb]
  \centering
  \begin{tikzpicture}
    \draw[fermion] (0,-0.293) node [left] {$t$} -- (1.414,-0.293);
    \draw[fermion] (1.414,-0.293) -- (1.414,-1.707);
    \draw[fermion] (1.414,-1.707) -- (0,-1.707) node [left]
    {$\bar{t}$};
    \draw[scalar] (1.414,-0.293) -- (2.828,-0.293) node[right] {$\rho$};
    \draw[gauge-na] (1.414,-1.707) -- (2.828,-1.707) node[right]
    {$g$};
    \draw[gauge-na] (4,0) node[left] {$g$}-- (5,-1);
    \draw[fermion] (4,-2) node[left] {$t$} -- (5,-1);
    \draw[fermion] (5,-1) -- (6.414,-1);
    \draw[fermion] (6.414,-1) -- (7.414,-2) node[right] {$t$};
    \draw[scalar] (6.414,-1) -- (7.414,0) node[right] {$\rho$};
  \end{tikzpicture}
  \caption{Dominant diagrams for the production of dark Higgs bosons.
  Both processes have an additional $u$-channel process that is not illustrated.}
  \label{fig:dh-prod}
\end{figure}
\noindent
The determination of the dark Higgs abundance involves the $2 \to 2$
processes of top quark annihilation and inelastic scattering as shown
in Figure \ref{fig:dh-prod}.  These processes freeze-in at temperatures
$T \lesssim m_t$.  Assuming that $m_\rho \ll m_t$, the dark Higgs mass can
be neglected.  These processes only turn on after the electrweak phase
transition, but nevertheless dominate production.  The total
cross-sections for these processes are then given by
\begin{equation}
  \label{eq:5}
  \sigma(t \bar{t} \to \rho g) = \frac{2\pi \epsilon^2 \alpha_s \alpha
    m_t^2 [\sqrt{s (s - 4 m_t^2)} (s^2 -12 m_t^2 s + 8 m_t^4) - 2 (s^3
    - 7 s^2 m_t^2 + 8 m_t^4 s - 16 m_t^6) {\rm arctan} \sqrt{1 - 4
      m_t^2 /s}]}{9 m_W^2 (m_t^2 - s) s^2 (s - 4 m_t^2) s_W^2}
\end{equation}
and
\begin{equation}
\label{eq:6}
  \sigma(t g \to t \rho) = \frac{\pi \epsilon^2 \alpha_s \alpha m_t^2
   [2 s^2 (3 m_t^2 + s)^2 \log(m_t^2 / s) - (s - m_t^2 ) (m_t^2 +3
    s)(m_t^4 - 8m_t^2 s -s^2)]}{24 m_W^2 s^2 s_W^2 (s - m_t^2)^3}
\end{equation}
respectively.  Thermally averaging\footnote{In doing so, we neglect backreaction terms for
  stimulated emission and fermi blocking.  Such terms are expected to
  have at most an $\mathcal{O}(1)$ effect in the relevant regime.} and
solving the Boltzmann equation numerically, we find
\begin{equation}
  \label{eq:7}
  Y(t \to \infty) \approx (0.27 + 0.42) \frac{\epsilon^2 \alpha
    \alpha_s m_\rho^5 m_t^2}{8 \pi^3 H(m_\rho) s(m_\rho) m_W^2 s_W^2},
\end{equation}
where the first number denotes the contribution from annihilation and
the second denotes the contribution from inelastic scattering of $t$
and $\bar{t}$.  Note that this result depends only on the mixing angle
$\epsilon$ in the limit that $m_\rho \ll m_t$.  Numerically, we find
\begin{equation}
  \label{eq:8}
  Y(t \to \infty) \approx 1.6 \times 10^{12} \times \epsilon^2
\end{equation}

\section{Mediator Decays}
\label{sec:mediator-decays}
Once the universe has cooled such that $\Gamma \sim H$, the mediator
will decay rapidly.  The consequences of this decay depend both on
when the decay occurs and on wha the dominant decay products are.
The Hubble rate during the period of BBN and during recombination are
well-known within standard cosmology, so given a determination of the
decay rate, we can determine the parameter space that can potentially
be covered by the analysis in this work.  We now present our
determination of both the total decay width and the branching
fractions of the mediators.

\subsection{Dark Photon Decays}
\label{sec:dark-photon-decays}

The dark photon decays to any charged particles with $m < m_{\gamma'} / 2$.
Decays to color neutral particles, which is to say leptons in the
part of parameter space considered here, are straightforwardly
determined using \eqref{eq:dp-decay}.  The hadronic width and branching
fractions are somewhat more involved.  Since the dark photon couples
to the electromagnetic current, hadronic decays can be related to the hadron-to-muon
cross-section ratio in $e^+ e^-$ interactions via
\begin{equation}
  \label{eq:10}
  \Gamma(\gamma' \to {\rm hadrons}) = R(E_{\rm CM} = m_{\gamma'})  \Gamma(\gamma' \to
  \mu^+ \mu^-).
\end{equation}
For $m_{\gamma'} \gtrsim 2~{\rm GeV}$, the ratio $R$ can be accurately
determined in perturbative QCD via
\begin{equation}
  \label{eq:11}
  R(m_{\gamma'}) = 3 \sum_f Q_f^2 \frac{(m_{\gamma'}^2 + 2 m_f^2) \sqrt{m_{\gamma'}^2 - 4
      m_f^2}}{(m_{\gamma'}^2 + 2 m_\mu^2) \sqrt{m_{\gamma'}^2 - 4 m_\mu^2}} \left(1  +
    \frac{\alpha_s}{\pi} + \mathcal{O}(\alpha_s^2)\right).
\end{equation}
The exclusive number of each type of quasi-stable hadron has been
determined using \verb+PYTHIA 6+ \cite{Sjostrand:2006za} to simulate a parton shower and
hadronization in $e^+ e^-$ collisions at $E_{\rm CM} = m_{\gamma'}$.

For $m_{\gamma'}\lesssim 2~{\rm GeV}$, we use data-driven methods to
determine both $R$ and the fragmentation into exclusive final states.
The ratio has been determined by summing the various exclusive final
states in several experiments at low energies and a combination of
these has been presented by the Particle Data Group \cite{Ezhela:2003pp,Agashe:2014kda}.  We then
determine the fragmentation into quasi-stable hadrons using the measured
branching fractions of the few resonances that contribute to $R$ at
low energies.

\begin{figure}[!tb]
  \centering
  \includegraphics[width=0.49\textwidth]{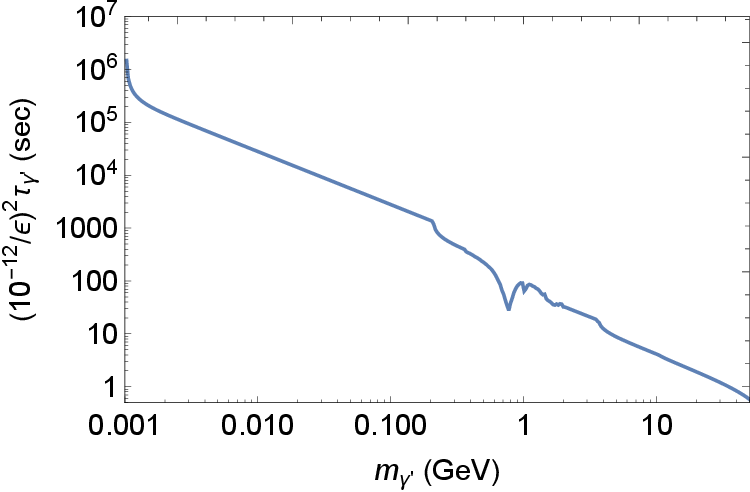}
  \includegraphics[width=0.49\textwidth]{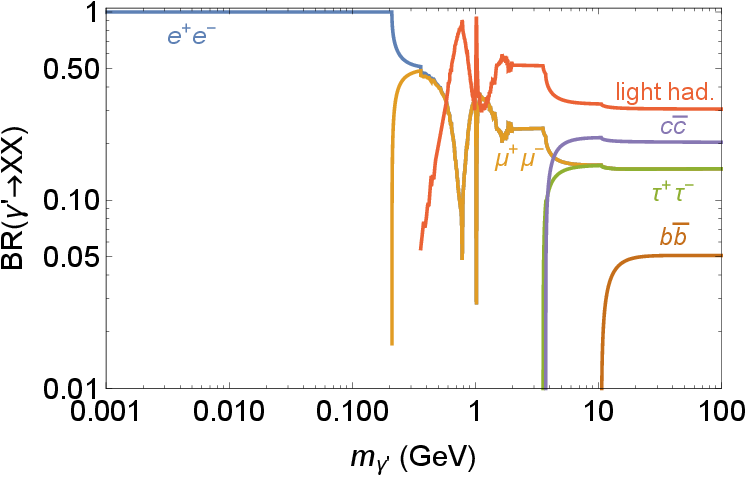}
  \caption{Lifetime and branching fraction of a dark photon. The 
lifetime becomes short when resonant hadronic decay occurs, as for example 
at $\sim 750\,$MeV, the
approximate mass of the $\omega$-resonance.}
  \label{fig:dp-decay}
\end{figure}
The resulting total decay width and branching fractions are shown in
Figure \ref{fig:dp-decay}.  

\subsection{Dark Higgs Decays}
\label{sec:dark-higgs-decays}

The dark Higgs decays with couplings that are proportional to those of
the SM Higgs.  For $m_\rho \gtrsim 2~{\rm GeV}$, we once more turn to
a perturbative determination of the dark Higgs decay width and
inclusive branching fractions.  Unlike in the dark photon case, decays
to pairs of gauge bosons (namely gluons and photons) are allowed and
can be significant in certain parts of parameter space.  The partial widths to
fermions are deterimed at leading order by
\begin{equation}
  \label{eq:dh-decay}
  \Gamma(\rho \to f \bar{f}) = \sin^2\epsilon \frac{G_f m_f^2}{4
    \sqrt{2} \pi} m_\rho \left(1 - \frac{4 m_f^2}{m_\rho^2}\right)^{3/2}
\end{equation}
For decays to quarks, an NLO correction factor of \cite{Braaten:1980yq}
\begin{equation}
  \label{eq:12}
  1 + 5.67 \frac{\alpha_s}{\pi} + \mathcal{O}(\alpha_s^2)
\end{equation}
is applied.  The decays to gluons and photons  (including
a NLO correction for the gluon case \cite{Inami:1982xt}) are given by
\begin{equation}
  \label{eq:13}
  \Gamma(\rho \to g g) = \sin^2 \epsilon \frac{G_f \alpha_s^2
    m_\rho^3}{64 \sqrt{2} \pi^3} \left|\sum_q F_{1/2}(\tau_q)\right|^2 \left(1 +
    \frac{215}{12} \frac{\alpha_s}{\pi} + \mathcal{O}(\alpha_s^2)\right)
\end{equation}
and
\begin{equation}
  \label{eq:15}
  \Gamma(\rho \to \gamma \gamma) = \sin^2 \epsilon \frac{G_f \alpha^2
    m_\rho^3}{128 \sqrt{2} \pi^3} \left|\sum_f N_{c,f} Q_f^2 F_{1/2}(\tau_f) + F_1(\tau_W)\right|^2 \left(1 +
    \frac{215}{12} \frac{\alpha_s}{\pi} + \mathcal{O}(\alpha_s^2)\right)
\end{equation}
respectively, with $\tau_i \equiv 4 m_i^2 / m_\rho^2$.  
$F_i$ are well-known loop functions, given explicitly in \cite{Gunion:1989we} for
example.  The exclusive fragmentation into quasi-stable hadrons is
once again determined using \verb+PYTHIA 6+ \cite{Sjostrand:2006za}.
In this case, we simulate $e^+ e^- \to h$ production at $E_{\rm CM} = m_\rho$.

For  $m_\rho \lesssim 2~{\rm GeV}$, there is no purely data-driven
method for determining the decay width and branching fractions as the
SM Higgs is far too heavy and weakly coupled to be seen in low-energy
$e^+e^-$ collisions.  There is some degree of controversy surrounding
such decays, with several methods presenting vastly different
results.  In this work, for masses below $1~{\rm GeV}$, we use a
determination based on low energy theorems, while we continue
to use a perturbative calculation down to around $1~{\rm GeV}$, where
the two methods overlap.  It is worth noting that light scalar
resonances can distort these results.

\begin{figure}[!tb]
  \centering
  \includegraphics[width=0.49\textwidth]{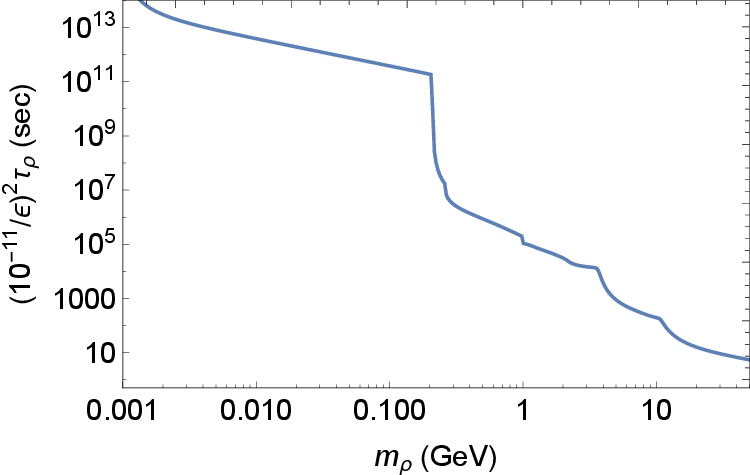}
  \includegraphics[width=0.49\textwidth]{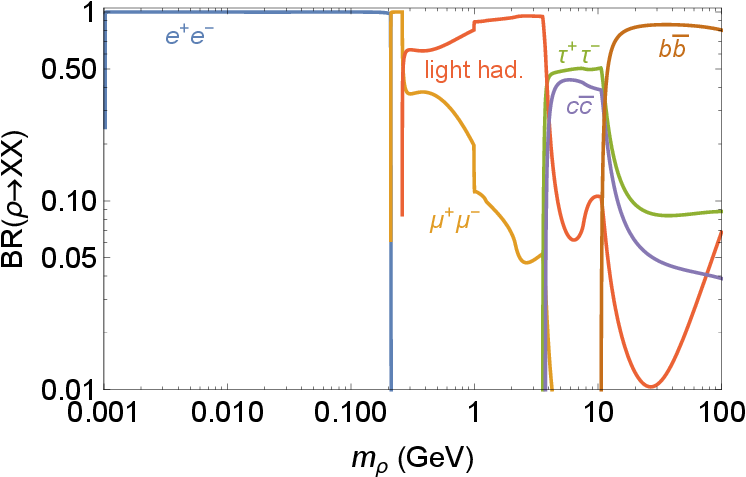}
  \caption{Lifetime and branching fraction of a dark Higgs.
  Note that the branching ratio to two photons is eceedingly small 
  and therefore not seen in the figure.}
  \label{fig:dh-decay}
\end{figure}
Low energy theorems predict decay widths of \cite{Vainshtein:1980ea,Voloshin:1985tc}
\begin{equation}
  \label{eq:16}
  \Gamma(\rho \to M \bar{M}) = \frac{1}{81} \frac{m_\rho^2}{m_\mu^2}
  \left(1 + \frac{11}{2} \frac{m_M^2}{m_\rho^2}\right)^2 \frac{(1
      - 4 m_M^2 / m_\rho^2)^{1/2}}{(1 - 4 m_\mu^2 / m_\rho^2)^{3/2}}
    \Gamma(\rho \to \mu^+ \mu^-),
\end{equation}
where $M \bar{M}$ are all isospin combinations (2 $\pi^+ \pi^-$,
$\pi^0 \pi^0$, 2 $K \bar{K}$,  and 2 $K^+ K^-$ being the relevant ones
for our study---$\eta$ is too heavy to contribute before we switch to a
perturbative calculation).  The resulting decay width and branching
fractions are shown in Figure \ref{fig:dh-decay}.

\section{Nucleosynthesis Constraints}
\label{sec:bbn}

It is well known that quasi-stable particles with decay times $\tau
\gsim 0.1\,$seconds may significantly perturb the primordial light
element nucleosynthesis occuring approximately between $1$ and
$1000$ seconds after the birth of the universe~\cite{Agashe:2014kda}.
Comparing the observationally inferred primordial abundances with the
predicted ones, we are able to derive limits on the abundance and
lifetimes of these putative relic particles.  In the previous
sections, we have derived the frozen-in abundances for dark Higgses
and dark photons for a generic model.  In this section, we consider
the constraints on the model parameter space from Big Bang
nucleosynthesis (BBN). 
\newline
\newline
When the quasi-stable mediators decay, their %
energetic decay products can perturb BBN by either hadronic or 
electromagnetic interactions with the particles in the baryon-photon plasma.  
In particular, the injection of mesons between $\tau \sim 0.1$-$10$ seconds 
may alter the neutron-to-proton ratio via charge exchange reactions, and 
thereby elevate the primordial helium mass fraction $Y_p$ beyond its 
observational upper limit.  The injection of energetic nucleons at 
$\tau > 100$ seconds produces a cascade of secondary and tertiary 
energetic nucleons which are capable of spalling $^4$He, thereby 
producing neutrons, $^2$H, $^3$H, and $^3$He. The resulting neutrons 
may form $^2$H via non-thermal fusion of protons.  Energetic $^3$H and 
$^3$He may fuse on $^4$He to form $^6$Li to generate an abundance orders 
of magnitude larger than what predicted in standard BBN.  Injection of 
energetic electromagnetically interacting particles, on the other hand, 
produce a cascade on the cosmic microwave background (CMB) radiation, 
with the resulting gamma-rays capable of photo-disintegrating $^2$H for 
$\tau \gsim 10^5$ seconds and $^4$He for $\tau\gsim 3\times 10^6$ seconds.  
Altogether, there are $\mathcal{O}(100)$ hadronic and electromagnetic 
interactions that are important.  
See~\cite{Jedamzik:2006xz} for additional details. 
\begin{figure}
\epsfxsize=12.5cm
\epsffile[50 50 410 302]{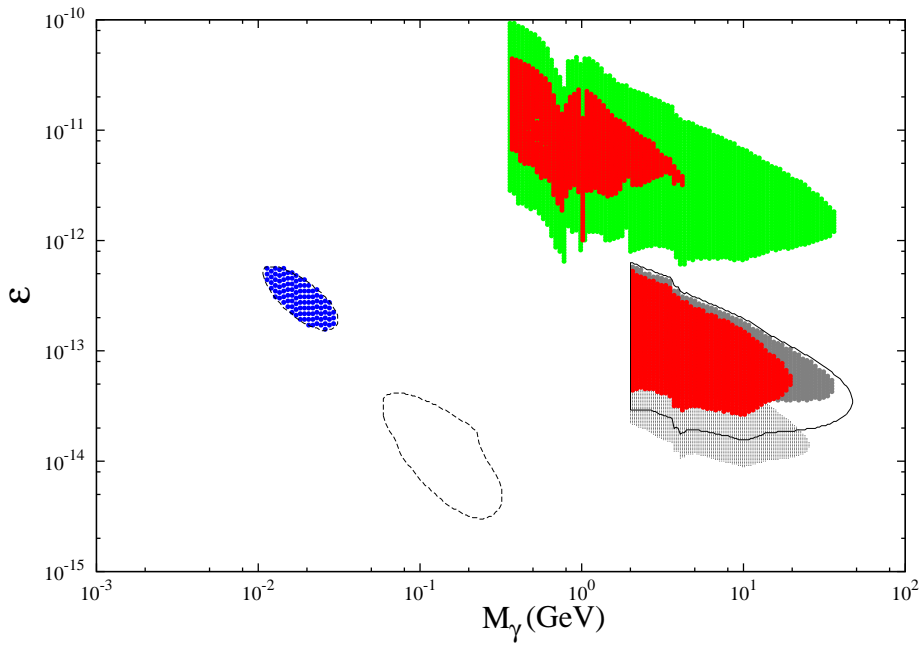}
\caption{Constraints on the dark photon mixing $\epsilon$-mass 
$m_{\gamma'}$ parameter space from Big Bang nucleosynthesis. 
Colored shaded regions are ruled out.  Shaded green and red areas are 
ruled out from $^4$He and $^2$H overproduction, respectively. In shaded 
blue areas $^2$H is underproduced.  In the light grey areas a substantial 
$^6$Li abundance is produced (i.e. $^6$Li/$^7$Li $> 0.1$), whereas in the 
dark shaded areas the $^7$Li abundance is reduced to a observationally favored 
$^7$Li/H $<
2.5\times 10^{-10}$. The solid line shows how much the red area ruled
out due to $^2$H overproduction would increase if a more stringent,
less conservative, $^2$H/H $< 3\times 10^{-5}$ (versus $4\times
10^{-5}$ limit) would be imposed. This indicates that solutions to the
cosmological $^7$Li problem only exist in models which predict $^2$H/H
$> 3\times 10^{-5}$.  Finally, in the region within the dotted line a
mild increase of the $^3$He/$^2$H ratio occurs. The ratio nevertheless
does not surpass either one or the observational limit of $1.5$.}
\label{fig2dp}
\end{figure}
\newline
\newline
In the following analysis, we adopt the following observational inferred 
constraints on primordial light-element abundances:
\begin{equation}
^2{\rm H/H} < 4\times 10^{-5}\, 
\end{equation}
from quasar absorption systems at high 
redshift~\cite{Burles:1997ez,Burles:1997fa,Cooke:2013cba}.  In addition,
\begin{equation}
^2{\rm H/H} > 2\times 10^{-5}\, ,
\end{equation}
from deuterium abundances in the local interstellar 
medium~\cite{Linsky:2006mr}.  
Here it is noted that a recent determination of the primordial deuterium
abundance of $2.53\pm 0.04\times 10^{-5}$~\cite{Cooke:2013cba} in a number
of damped Lyman-$\alpha$ systems
would substantially increase the parameter space
which is ruled out. However, we do not use these very stringent limits
as it has not been established that prior determinations of $^2$H/H in lower
column density quasar absorption systems going to values as 
large as $5.3\times 10^{-5}$~\cite{Burles:1997ez,Burles:1997fa}
(at two sigma) may be flawed. We feel that the recent determination could
actually be biased to demonstrate concordance between the standard BBN
prediction at a precisely inferred
baryon density from observations of the cosmic microwave background radiation
(hereafter, CMBR) 
and observations. Though this concordance generally exists, the presented new
analysis does not provide significant observational improvements to warrant 
such a large reduction of observational error bars.
It has been shown that $^3$He/$^2$H is an
important diagnostic~\cite{Sigl:1995kk} and we adopt
\begin{equation}
^3{\rm He/^2H} < 1.5\, , 
\end{equation}
from the deuterium and $^3$He abundance in the presolar 
nebula~\cite{GG}.
The upper limit on the helium mass fraction $Y_p$ is taken to be 
\begin{equation}
Y_p < 0.26\, , 
\end{equation}
from extragalactic low-metallicity HII regions and an extrapolation to zero 
metallicity~\cite{Izotov:2013waa}.
In our figures we also indicate regions where an important
$^6$Li production of $^6$Li/$^7$Li $>0.1$ occurs. However, though the
observational upper limit on $^6$Li in the atmospheres of low-metallicity
Pop II stars is approximately $^6$Li/$^7$Li $\lsim 0.05$, such regions should 
strictly speaking not be considered ruled out, as $^6$Li is very fragile and
may have been destroyed in these stars from an initially higher level. This
becomes particularly likely since $^7$Li in the atmospheres of such stars
is mostly observed in the range $1\times 10^{-10}\lsim {\rm ^7Li/H}\lsim
2.5\times 10^{-10}$~\cite{lithium} 
as opposed to the expected primordial prediction of
$^7$Li/H $\approx 5\times 10^{-10}$  in standard BBN. The leading astrophysical
explanation for this discrepancy is $^7$Li destruction in low-metallicity Pop II
stars due to an unknown mixing process in such stars. If this, indeed, is the
case, $^6$Li will be destroyed by even a larger factor than $^7$Li as it is
more fragile, and thus is not readily usable as constraint. However, an 
alternative explanation of the "cosmological lithium problem" is a destruction
of $^7$Li (often concommitant with production of $^6$Li)
by the decay of relic particles~\cite{Jedamzik:2004er}. This may also be 
achieved by decaying
dark photons or dark Higgs with the right parameters. We thus show regions of parameter space 
\begin{equation}
9\times 10^{-11}< \,^7{\rm Li/H} < 2.5\times 10^{-10}\, ,
\end{equation}
where the $^7$Li problem is supposed to be strongly alleviated or solved
by the decay of such particles. Finally, in a very small parameter space 
constraints apply from the potential underproduction of $^7$Li, we adopt
$^7$Li/H $> 9\times 10^{-11}$ as a conservative constraint.

\subsection{BBN Constraints on Dark Photons}

\begin{figure}
\epsfxsize=12.5cm
\epsffile[50 50 410 302]{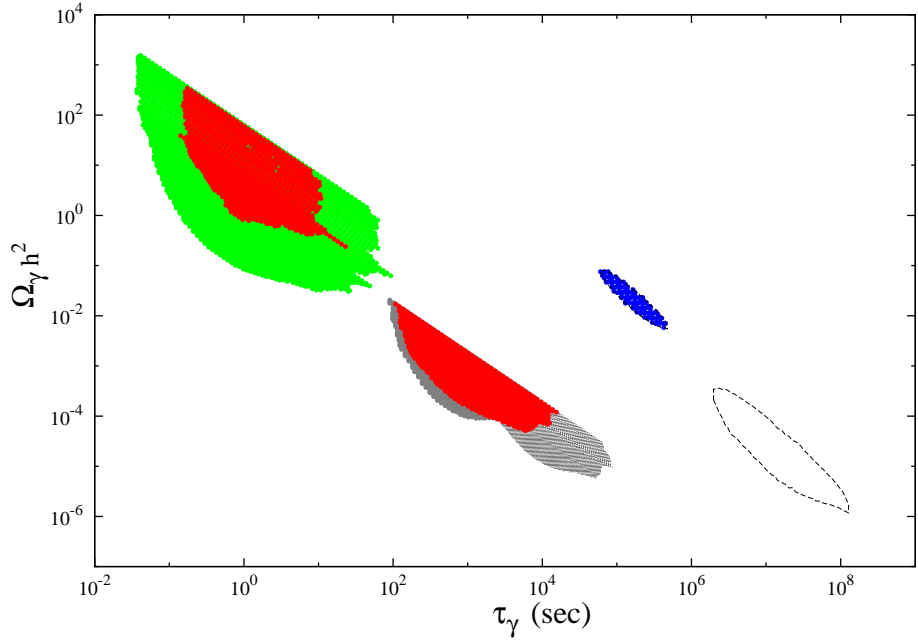}
\caption{The cosmological parameters associated with the
dark photon models in Fig.~\ref{fig2dp} with the same color coding as
in Fig.~\ref{fig2dp}. Shown are the dark photon lifetime and it's abundance
$\Omega_{\gamma'}h^2$,
in particular, the contribution the the present critical density dark photons
would have if they did not decay. Note that $h$ is the present
Hubble constant in units of 100 $\rm km\, s^{-1}\, Mpc^{-1}$.}
\label{fig3dp}
\end{figure}

The lifetime of the dark photon is shown in Fig.~\ref{fig:dp-decay}. It is seen that
for sufficiently small $\epsilon$ and $m_{\gamma'}$ it may be
much longer that $\tau\sim 1\,$sec. As the dark photon shares the quantum
numbers of the photon, decay occurs into charged particles, whenever
kinematically allowed. Note that decay into light quarks is kinematically
blocked already 
for $m_{\gamma'} < 2 m_{\pi^0}$ as otherwise hadronization may not occur.
When 
$m_{\gamma'}$ approaches $2m_e$ from above the lifetime increases
due to reduced phase space in the decay into $e^{\pm}$. For 
$m_{\gamma'} < 2m_e$ decay has to
occur over a loop diagram into three photons with the respective lifetime
becoming very long. This case is not treated in what follows.

In Fig.~\ref{fig2dp} the shaded colored regions show all the parameter space which
is ruled out from a comparison between predicted and observationally inferred
abuandances. The constraints are from $^4$He overproduction (green), $^2$H
overproduction (red), and $^2$H underproduction (blue). The light grey area
indicates parameter space where the cosmological $^7$Li problem could be 
alleviated/solved, wheres the grey area indicates significant $^6$Li production.
Also shown in this figure by the solid line, is shown of how much the $^2$H
overproduction region would grow if a more agressive limit of 
$^2$H/H $<3\times 10^{-5}$ would be imposed. These results thus
confirm the known
conclusion that solutions to the $^7$Li problem through the decay of massive 
relic particle
may not be attained without some additional $^2$H/H production. In particular,
no scenario solving the $^7$Li problem with decaying dark photons achieves 
$^2$H/H $<3\times 10^{-5}$. 

\begin{figure}
\epsfxsize=12.5cm
\epsffile[50 50 410 302]{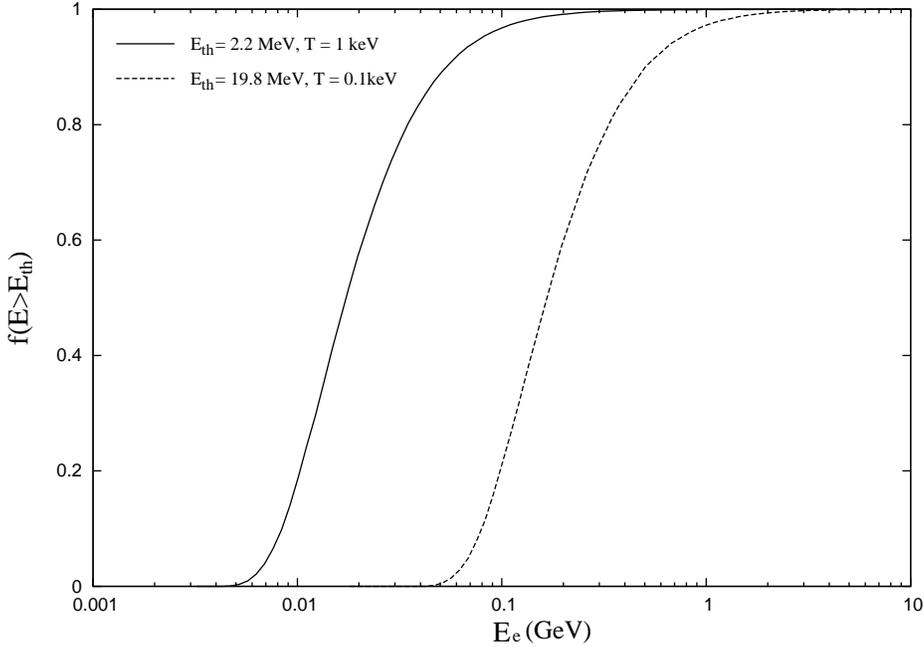}
\caption{
The fraction of energy converted into photons capable
of photodisintegration either $^2$H or $^4$He, 
produced when electron-positron pairs of energy $2E_e$ are injected.
Two examples are shown, (a) injection of pairs at cosmic temperature
$T = 1\,$keV and $E_{th} = 2.2\,$MeV for deuterium photodisintegration, 
and (b) $T = 0.1\,$keV and $E_{th} = 19.8\,$MeV for helium photodisintegration. 
It is seen that in both cases the electron (positron) energy $E_e$
has to be a factor ten larger in order to reach fractions of order one half.}
\label{fig4dp}
\end{figure}

In Fig.~\ref{fig3dp}, the predicted abundances $\Omega_{\gamma'}h^2$
and lifetimes $1/\Gamma_{\gamma'}$ for the models excluded in 
Fig.~\ref{fig2dp}
are shown with the same color coding. It is noted that well-known trends are
followed, i.e. relic particles hadronically decaying, (i.e. 
$m_{\gamma'} > 2m_{\pi}$ with shorter lifetimes $\tau\sim 1\,$sec but
higher abundances $\Omega_{\gamma'}h^2$ are mostly constrained by $^4$He
overproduction, hadronically decaying relics with $\tau\sim 1000\,$sec
and $\Omega h^2\sim 10^{-3}-10^{-4}$ are constrained by $^2$H
overproduction, and electromagnetically or hadronically decaying relics with
longer lifetimes $\tau \gsim 10^5\,$ sec may be constrained by $^2$H
underproduction. However, one usual trend is not observed in these figures.
For the typical massive decaying particle stringent constraints apply at
$\tau \gsim 3\times 10^6$ sec at already fairly
low $\Omega h^2$ due to $^4$He photodisintegration and the
concomitant $^3$He/$^2$H overproduction. This has also been claimed by
the recent analysis of Ref.~\cite{Fradette:2014sza} 
for the case of the dark photon. Usually 
injected energetic $e^{\pm}$ rapidly inverse Compton scatter on the CMBR, with 
the
resultant $\gamma$-rays pair-producing $e^{\pm}$ pairs on the CMBR, leading to
a new lower energy generation of $e^{\pm}$. The process repeats itself until
the resulting $\gamma$-rays have energy 
$E_{\gamma} \lsim E_{\gamma}^{e^{\pm}}\approx m_e^2/(20T)$ and are no
longer sufficiently energetic to further 
pair-produce on thermal CMBR photons. When the cosmic temperature is low
enough, such $\gamma$-rays may then photodisintegrate $^2$H 
and $^4$He with photodisintegration thresholds $E_{\gamma}^{th}=\,$
2.2 MeV and 19.8 MeV,
respectively. Such a scenario has been erroneously assumed by 
Ref.~\cite{Fradette:2014sza}.
However, for the particular $\tau_{\gamma'}$ - $m_{\gamma'}$
relation imposed by the dark photon interactions, the energy of
injected $e^{\pm}$ due to dark photon decay is often below
$E_{\gamma}^{e^{\pm}}$. Moreover,
for most of the parameter space the injected energies of $e^{\pm}$ are so low
that they are well in the Thomson scattering regime, i.e $E_e\ll m_e^2/T$
(as opposed to the relativistic Klein-Nishima scattering regime).  It is well 
known that
inverse Compton scattering of $e^{\pm}$ with energy $E_e$ in the Thomson scattering 
regime leads to a multidude
of softer photons with $E_{\gamma}\ll E_e$, rather than to a smaller number of 
energetic $\gamma$-rays with $E_{\gamma}\sim E_e$. This is exemplified by an
accurate calculation shown in Fig.~\ref{fig4dp}, which shows the fraction of produced
$\gamma$-rays capable of photodisintegrating $^2$H and $T=1\,$keV of $^4$He
at $T=0.1\,$keV as a function of dark photon mass. It is seen that for 
$m_{\gamma'}$  too small, this fraction becomes very small, though 
naively one would expect photodisintegration to be possible as 
$m_{\gamma'}\gg E_{\gamma}^{th}$. The region enclosed by the
dashed line in Fig.~\ref{fig2dp} and Fig.~\ref{fig3dp} thus shows parameter 
space which lead to a
slightly elevated $^3$He/$^2$H ratio, nevertheless, far from being ruled
out.


 
\subsection{BBN Constraints on Dark Higgses}

Decay times of dark Higgs for fixed mixing angle span a much larger range
than those of dark photons. This may be seen in Fig.~\ref{fig:dh-decay}. 
The dark Higgs has the same interactions with standard model particles
as the Higgs particle up to an overall factor of $\epsilon$. As such,
its lifetime scales $1/(\epsilon^2m^2)$, where $m$ is 
the mass of the standard model particles in arising from the dominant decay.
Thus, whenever a channel becomes kinematically forbidden at
$m_{\rho}<2m_1$, decay predominantly occurs to the heaviest of all the
lighter particles where decay is not kinematically forbidden. If this particle
has mass $m_2$, the decay time increases essentially instantaneously by
$(m_1/m_2)^2$. This trend can be seen in Fig.~\ref{fig:dh-decay},
where sudden increases of $\tau_{\rho}$ are observed, as for example at
$m_{\rho}\approx 2 m_{\pi}$ where predominant decay switches first from
hadrons to $\mu^{\pm}$ and than quickly to $e^{\pm}$ yielding an about five
order of magnitude increase in the decay time. The arguments above of course
do not apply to loop-diagram decay into photons ($\gamma\gamma$) and gluons
$(gg)$. However, these decay channels are only important in very narrow intervals,
as for example seen in the branching ratio in Fig.~\ref{fig:dh-decay}, 
which shows that decay
$\rho\to \gamma\gamma$ dominates only when $m_{\rho}$ comes too close
to the $\rho\to e e^+$ threshold. For such low $m_{\rho}$ dark Higgs
decay may occur with lifetimes comparable or longer than the lifetime of the 
current
Universe. Treating the important $x$- and $\gamma$-ray constraint which should
apply in this regime is beyond the scope of the current paper, such that our
CMBR analysis below only treats the regime $m_{\rho}> 2\,$MeV.   

\begin{figure}
\epsfxsize=12.5cm
\epsffile[50 50 410 302]{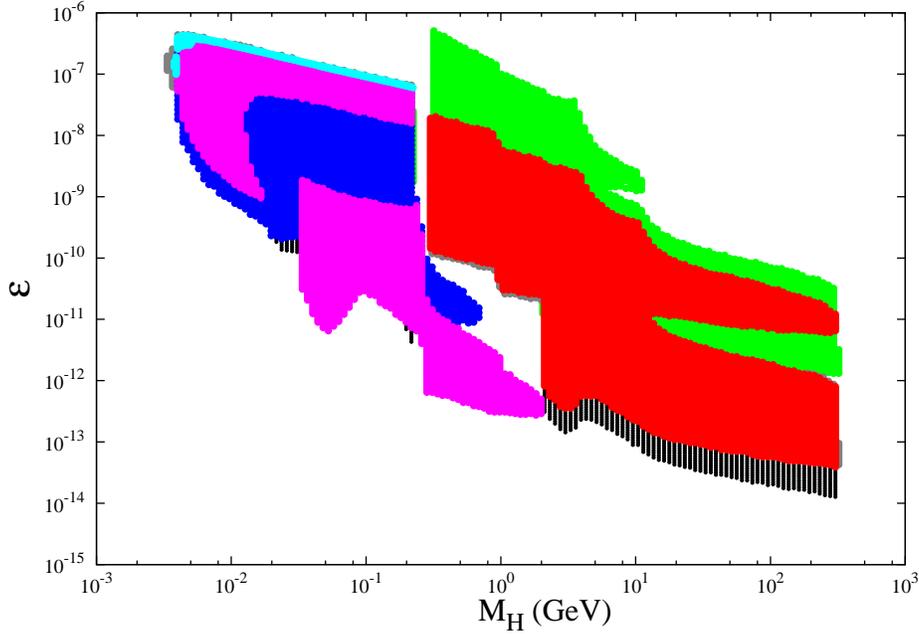}
\caption{The constraints on the dark Higgs parameter space with the
color/line coding as in Fig.~\ref{fig3dp}. Solid purple areas are ruled
out due to an overproduction of the $^3$He/$^2$D ratio, i.e. $^3$He/$^2$D
$> 1.5$.}
\label{fig3dh}
\end{figure}

\begin{figure}
\epsfxsize=12.5cm
\epsffile[50 50 410 302]{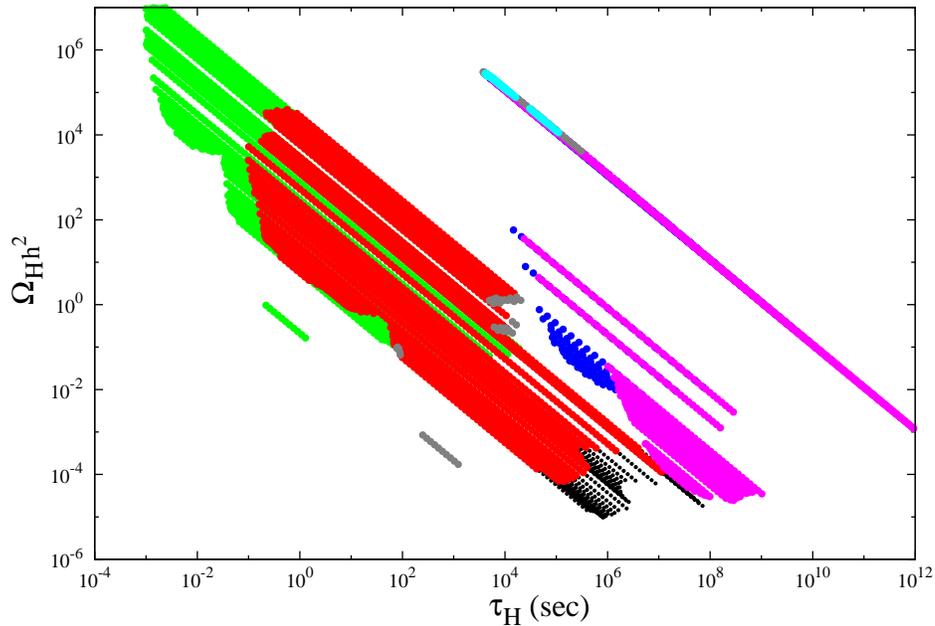}
\caption{The Higgs abundances and lifetimes associated with the dark Higgs
parameter space ruled out as shown in Fig.~\ref{fig3dh}.}
\label{fig4dh}
\end{figure}

In Fig.~\ref{fig3dh} BBN constraints on dark Higgses in the mass-mixing angle plane
are shown. The color coding is as in Fig.~\ref{fig2dp}, except for the purple regions
which are ruled out by $^3$He/$^2$H overproduction and the very small 
light-blue regions which are ruled out by $^7$Li underproduction. Though
similar arguments as for the dark photon, concerning the efficiency of
$^4$He photodisintegration in the Thomson scattering regime apply to the
dark Higgs, since their decay channel below $m_{\rho} < 2 m_{\pi}$ is
also into electrons, large areas are ruled out from $^3$He/$^2$H overproduction
simply because the abundance of dark Higgses is typically larger than that
of dark photons. This may be seen from Fig.~\ref{fig4dh} which shows the ruled out 
$\tau_{\rho}$-$\Omega_{\rho}$ plane for the same models as in 
Fig.~\ref{fig3dh}.
Some of the possible $\tau_{\rho}$-$\Omega_{\rho}$ 
parameter space is not covered, as the constraints come on seperated lines
in the $\tau_{\rho}$-$\Omega_{\rho}$ plane. This is simply an artifact
due to our grid spacing of $1.06$ in $m_{\rho}$ (as well as in $\epsilon$.
For jumps in $\tau_{\rho}$ as large as $10^5$ over a narrow $m_{\rho}$
interval not all possible decay times are resolved, leading to this line
structure, with individual neigboring lines seperated by in mass by a factor 
$1.06$.

We note that the $^7$Li problem can be solved at $\epsilon \sim
10^{-7}$ and $m_{\rho}\approx 3.5\,$MeV. This portion of parameter space is close to 
regions which are ruled out by $^7$Li underproduction, i.e.\ $^7$Li 
$<9\times 10^{-11}$.  We call attention to this particular 
region, even though it is small, as it is the only currently known solution to the $^7$Li problem by
decay of relic particles which does ($\it not$) lead to additional $^2$H 
production or destruction. The reason for this is simple; the injection of
$e^{\pm}$ with energy $1.75\, {\rm GeV}$ may lead to $\gamma$-rays of energy which
are capable pf photodisintegrating $^7$Be with photodisintegration
threshold of $1.58\,$MeV but not $^2$H with threshold $2.2\,$MeV. As at 
earlier times most of the primordial $^7$Li is still in form of $^7$Be before
electron-capturing to form $^7$Li after 53 days, 
this becomes possible. Most other $^7$Li
solving scenarios rely on injection of neutrons which lead to an additional
$^2$H production via neutron capture on protons.

However, it is currently not clear if these already small $^7$Li solving 
regions  
do not become even smaller by $^4$He overproduction. We have noted that the 
typical abundances of dark Higgs are large. In fact they are so large that
they may come close to their thermal equilibrium abundance in some part of 
parameter space at very large $\epsilon$. It is well known that substantial extra
energy density over that of the standard model plasma, present at the onset
of BBN at $\tau\sim 1\,$sec, leads to an increase of the $^4$He abundance
due to an increase in the cosmic expansion rate. Such limits are often 
stated in terms of extra putative neutrino (light) degrees of freedom 
present in the plasma, with $\Delta N_{\nu}$ constrained to be smaller than
$\sim 1-2$. Note that the energy density $\Delta N_{\nu}$ at $T \sim 1\,$MeV
corresponds very approximately to $\Omega_{\rho}h^2\sim 3\times 10^{4}$ in dark
Higgs abundance. Such large values of $\Omega_{\rho}h^2$ with lifetimes
$\tau \gsim 1\,$sec are reached in the low-mass, high mixing angle region,
approximately for $\epsilon\gsim 10^{-6}-10^{-7}$ and 
$2\, {\rm MeV} \lsim m_{\rho} \lsim 300\,$MeV. Of course, dark Higgs
of $m_{\rho} > 1\,$MeV are not exactly light degrees of freedom 
at $T\lsim 1\,$MeV. Their energy density redshifts differently than that of
light degrees of freedom. At such large interaction strength it may also be
possible that dark Higgs will again be destroyed before BBN by process such as 
$\gamma \rho\to e^- e^+$.
All these effects are not included in the current analysis as they are beyond
the scope of the paper. In particular, BBN codes with decaying or annihilating
particles do usually not include such effects, as constraints are already 
usually stringent for much smaller abundances. We only note here, that 
the very upper left-hand corner of Fig.~\ref{fig3dh} is 
tagged to be potentially ruled out by $^4$He overproduction
and leave exact results for future work.

\section{Cosmic Microwave Background Constraints}
\label{sec:cmb}

\begin{figure}
\epsfxsize=12.5cm
\epsffile[50 50 410 302]{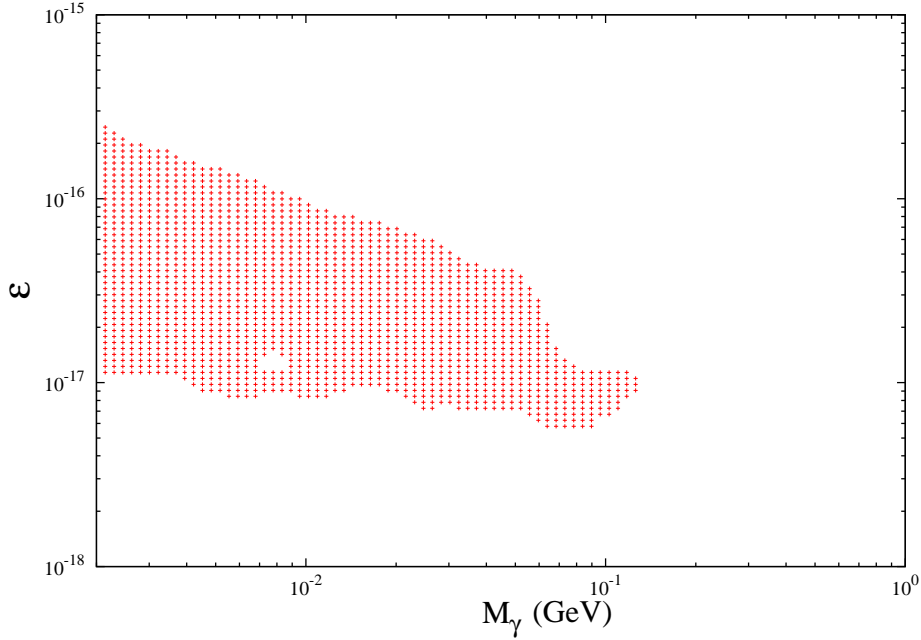}
\caption{The shaded region indicates dark photon models which are ruled
out by current observations of anisotropies in the CMBR.}
\label{fig1dp_cmb}
\end{figure}

\begin{figure}
\epsfxsize=12.5cm
\epsffile[50 50 410 302]{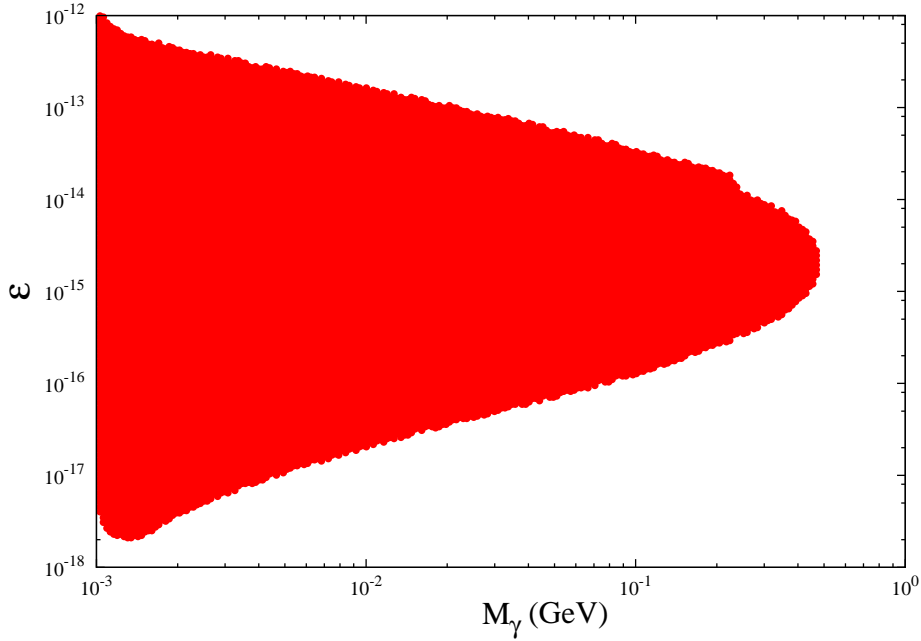}
\caption{Dark photon model which could be ruled out by a future
PIXIE mission.}
\label{fig2dp_cmb}
\end{figure}

Precise observations of the CMBR have proven to be an invaluable tool 
for determining cosmological parameters, as well as constraining
models of the early Universe beyond the standard model.
Observations of anisotropies in the CMBR, as well
as limits on deviations from a perfect blackbody spectrum of the CMBR,
have to be confronted by theoretical predictions for any particular model.
The Planck and WMAP satellite missions, as well as a large number of
ballone-type experiments, have observed anisotropies in the CMBR, whereas
the FIRAS-instrument on the COBE sattelite has limited deviations from an
ideal
Planck spectrum, to be approximately below one part in 
$10^4$~\cite{Fixsen:1996nj}. The latter
limits may be potentially improved by 3 to 4 orders of magnitude by a 
dedicated satellite mission named PIXIE \cite{Kogut:2011xw}.

The effects of annihilation and decay of relic particles on the anisotropies
of the CMBR have already been analyzed in a number of prior publications.  See for example~\cite{Finkbeiner:2011dx} and the references therein.  
Injecting energetic photons and/or electrons
into the plasma starts a cascade on the CMBR, producing a multitude of seconday
lower energy electrons and photons. When their energy falls below 
$E\sim 10\,$keV their main effect is the ionization and accompanying heating
of neutral hydrogen and helium.  
If the decay happens shortly before, during, or somewhat after recombination, 
an additional (scale-dependent) supression of the primary anisotropies by additional scattering of
CMBR photons on free electrons results. Though this effect may be partially
compensated for by increasing the normalization $A_s$ and the spectral index 
$n_s$ of the scalar primordial perturbations, the precise observations of the
high-l $TT$ anisotropy spectrum still results in important constraints.

In this study we have modified the recombination
routine RECFAST~\cite{Seager:1999bc} 
to include energy injection, employing the detailed
cascade results of Ref.~\cite{Slatyer:2012yq}. 
We assume that $1/3$ of the energy of
$E \lsim 10\,$keV $e^{-}$'s and $\gamma$'s goes into ionization, 
whereas $2/3$ into heating. The modified recombination routine was then used
with the CMBR code CAMB and the Markov-Chain Monte-Carlo (i.e. MCMC)
code COSMOMC~\cite{Lewis:2002ah} to derive constraints on dark photons and dark Higgses 
decaying around recombination. In our MCMC analysis we take very generous 
priors on cosmological parameters, in particular generous on $n_s$ and $A_s$,
to avoid missing degenerate good fits to the data. Results are produced 
typically by calculating of the order $2\times 10^6$ to $5\times 10^6$ models, 
and calculating their liklihood that they may produce the observed data.

It is well known that the injection of energetic $e^{\pm}$'s and $\gamma$'s
before recombination (as well as after in the case of $e^{\pm}$) leads to
spectral distortions in the CMBR energy spectrum, since the energy injected
may not be anymore entirely thermalized when injected below redshift 
$z\lsim 3\times 10^6$ (temperature $T\lsim 1\,$keV). This leads to 
non-vanishing chemical potential $\mu$ deviations for 
$z \lsim 3\times 10^6$, since photon number changing
double Compton-scattering is no longer
efficient at these lower redshifts to produce the correct number of photons
in the spectrum, and spectral $y$ distortions for $z\lsim 4\times 10^4$ as even
Thomson scattering may then no longer completely equilibrate the distribution.
Such limits are generic, and may be stated as an upper limit on
the fractional energy $\Delta\epsilon^{\rm inj}/\epsilon^{\rm CMBR}$
of the total CMBR energy density, which is due to
non-thermal injection. The FIRAS instrument imposes 
$\Delta\epsilon^{\rm inj}/\epsilon^{\rm CMBR}\lsim 6\times 10^{-5}$. A future
PIXIE mission would hope to reach the sensitivity of 
$\Delta\epsilon^{\rm inj}/\epsilon^{\rm CMBR}\lsim 10^{-8}$. However, it should
be noted that at such low $\Delta\epsilon^{\rm inj}/\epsilon^{\rm CMBR}\sim
10^{-8}$ other standard sources such as Silk-damping could become foregrounds.
In any case, such considerations set important limits already now, particularly
for the typical large predicted abundances of dark Higgs.

\subsection{CMB Constraints on Dark Photons}

Fig.~\ref{fig1dp_cmb} shows by the shaded area dark photon models in the
mass - mixing angle plane which are ruled out by the precise observations 
of the angular anisotropies of the CMBR. Here we used recent 
Planck data~\cite{Planck} for our analysis.
In general about as much parameter space is
ruled by the CMBR as by BBN, however, at lower masses and smaller mixing angles,
leading to decay around recombination. A small island of allowed models 
surrounded by disallowed models is noted in this figure. This is due to a
statistical fluctuation, since the entire ruled out region is not ruled out 
by a large likelihood degradation.
There are no constraints on dark photon parameters from current FIRAS
limits on deviations from a blackbody 
spectrum. In Fig.~\ref{fig2dp_cmb}
possible limits are shown after a sucessful PIXIE 
mission~\cite{Kogut:2011xw}. Here a maximal
sensitivity of $\Delta\epsilon^{\rm inj}/\epsilon^{\rm CMBR}\sim 10^{-8}$ 
has been assumed.
It is seen that such limits could rule out a multiple of the parameter space
currently ruled out by the anisotropies of the CMBR, of course, assuming that 
known foregrounds could be controlled.

\subsection{CMB Constraints on Dark Higgs}

\begin{figure}
\epsfxsize=12.5cm
\epsffile[50 50 410 302]{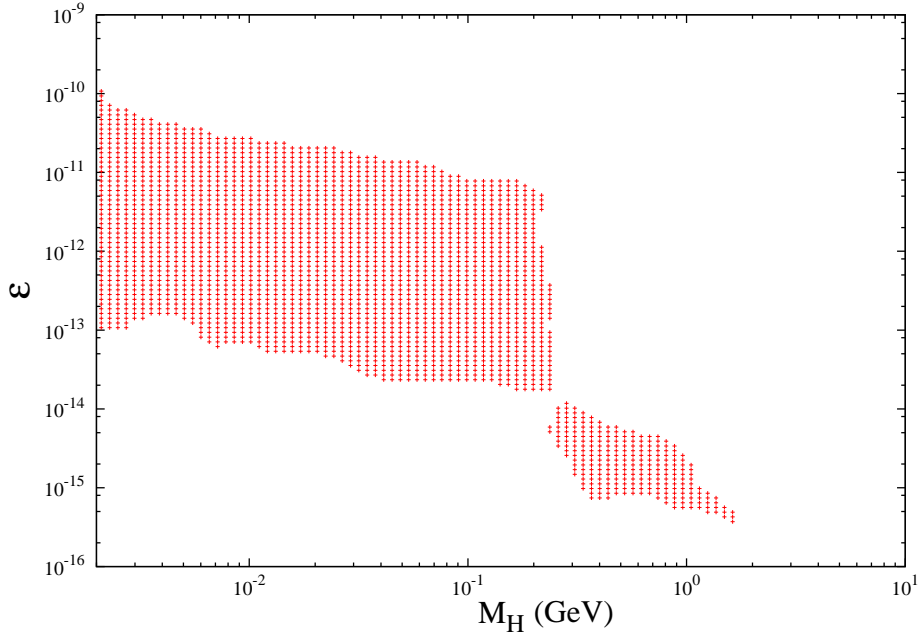}
\caption{Dark Higgs model ruled out by Planck data.}
\label{fig1dh_cmb}
\end{figure}

\begin{figure}
\epsfxsize=12.5cm
\epsffile[50 50 410 302]{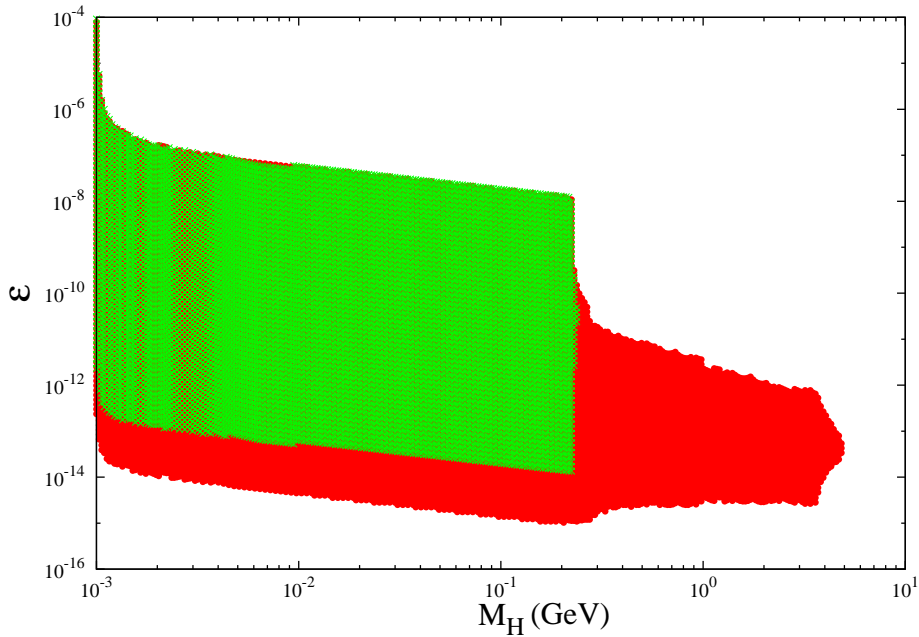}
\caption{Dark Higgs model currently ruled out by deviations of the CMBR
form a blackbody spectrum (green). The red area shows how much these limits
would extend after a PIXIE mission.}
\label{fig2dh_cmb}
\end{figure}

Limits on dark Higgs parameters from the anisotropies of the CMBR are shown
in Fig.~\ref{fig1dh_cmb}. Here as well somewhat disconnected pieces may be observed,
one at lower mass where the decay channel is $\rho\to e^-e^+$, and one
at somewhat higher masses where the main decay channel is 
$\rho\to {\rm hadrons}$, and where the $\rho$ lifetime is considerably 
smaller for the same $\epsilon$. In contrast to the case of the dark photon,
due to the high $\rho$ abundances, the ruled out regions show a very
significant degrade in the liklihood, i.e. are strongly ruled out. 
In Fig.~\ref{fig2dh_cmb} limits on the dark Higgs are shown from possible $\mu$
and $y$-type deviations of the Planck spectrum. Here the large green area
is ruled out by the current FIRAS limit, whereas it would only
somewhat extend to the red area with a sucessful PIXIE mission.




\section{Conclusions}
\label{sec:conclusions}

Bosonic mediators generically arise in models with a hidden dark
sector. In terms of models with renormalizable
interactions, the only two bosonic possibilities are a dark photon that
kinetically mixes with the standard model photon/$Z$ and a dark Higgs
that mixes with the standard model Higgs. Light mediators with masses
$m \lesssim 100\,$GeV are kinematically accessible in experiments, yet
can evade detection in accelerator and beam dump experiments if they
have extremely weak interactions with the standard model. When the
mixing becomes extremely weak the existence of such particles can be
tested by cosmology. In this work, we have performed a detailed study
of these tests. Assuming conservatively that their abundance is
negligible at the end of inflation they will nevertheless be produced in small abundance by two-body 
interactions in the thermal standard model plasma, with their 
non-equilibrium abundance governed by the small mixing parameter $\epsilon$, a
process dubbed freeze-in. Due to the smallness of $\epsilon$  they become
long-lived particles subsequently decaying back into the visible sector.
If this decay occurs after the onset of BBN, stringent constraints may be
derived from the distortion of the light-element nucleosynthesis, the 
distortion of the CMB blackbody spectrum, as well as the alteration of the
angular anisotropies in the CMB. 

We have found that large parts of parameter space at very small mixing  
$\epsilon\sim 10^{-6} - 10^{-14}$ are already constrained, or 
can be constrained by future missons such as PIXIE. Our results begin
the important task of constraining the small mixing angle regions of
models in which a mediator mixes with a standard model boson. We
further find that in very small parts of
parameter space such exotic particles could solve the cosmological lithium
problem. 

Several directions for further study are possible within the context
of these models.  For very light mediators, below around 1 MeV, new
decay modes and signals open up.  For dark Higgs, the dominant decay
is to a pair of photons.  For dark photons, decays to two
photons are forbidden by the Landau-Yang theorem.  Nevertheless,
loop-induced decays to three photons are possible. Either case could yield
interesting signals in the cosmological x-ray and gamma-ray
backgrounds. For large mixing angles, the mediator may thermalize with the standard
model before decaying.  Such a scenario has been mentioned in the text
above, but it would be interesting to verify the cosmological
constraints in detail. The addition of a stable dark matter particle
to the scenario discussed in this work could lead to an viable model
of either thermal or non-thermal cold dark matter production. The
simple models we have examined potentially have a very rich
phenomenology that remains to be explored further.

\vskip 0.2cm
\noindent
{\it Acknowledgments:}
We thank D.~Finkbeiner for collaborating during the early stages of
this project.  We also thank R.~Caldwell, D.~Curtin, S.~El~Hedri,
R.~Mohapatra, S.~Nussinov, and T.~Slatyer for useful conversations.
JB is supported by by the U. S. Department of Energy under the
contract DE-FG-02-95ER40896.  DW~is supported by a grant from the
University of Washington.  Part of this work was completed at SLAC,
which is operated by Stanford University for the 
US Department of Energy under contract DE-AC02-76SF00515.

\end{document}